\RequirePackage{lineno}

\documentclass[aps,prl,twocolumn,superscriptaddress,showpacs,preprintnumbers,amsmath,amssymb]{revtex4-1}

\usepackage{graphicx} 
\usepackage{dcolumn}  
\usepackage{color} %
\RequirePackage{xspace}
\usepackage{relsize}

\graphicspath{{ps}}

\begin{document}


\title{ \quad\\[1.0cm] 
Search for \mbox{\boldmath$\Lambda_c^+\to\phi p \pi^0$} and branching fraction measurement of  \mbox{\boldmath$\Lambda_c^+\to K^-\pi^+ p \pi^0$} }


\noaffiliation
\affiliation{University of the Basque Country UPV/EHU, 48080 Bilbao}
\affiliation{Beihang University, Beijing 100191}
\affiliation{Budker Institute of Nuclear Physics SB RAS, Novosibirsk 630090}
\affiliation{Faculty of Mathematics and Physics, Charles University, 121 16 Prague}
\affiliation{Chonnam National University, Kwangju 660-701}
\affiliation{University of Cincinnati, Cincinnati, Ohio 45221}
\affiliation{Deutsches Elektronen--Synchrotron, 22607 Hamburg}
\affiliation{University of Florida, Gainesville, Florida 32611}
\affiliation{Gifu University, Gifu 501-1193}
\affiliation{SOKENDAI (The Graduate University for Advanced Studies), Hayama 240-0193}
\affiliation{Gyeongsang National University, Chinju 660-701}
\affiliation{Hanyang University, Seoul 133-791}
\affiliation{University of Hawaii, Honolulu, Hawaii 96822}
\affiliation{High Energy Accelerator Research Organization (KEK), Tsukuba 305-0801}
\affiliation{J-PARC Branch, KEK Theory Center, High Energy Accelerator Research Organization (KEK), Tsukuba 305-0801}
\affiliation{IKERBASQUE, Basque Foundation for Science, 48013 Bilbao}
\affiliation{Indian Institute of Technology Bhubaneswar, Satya Nagar 751007}
\affiliation{Indian Institute of Technology Guwahati, Assam 781039}
\affiliation{Indian Institute of Technology Madras, Chennai 600036}
\affiliation{Indiana University, Bloomington, Indiana 47408}
\affiliation{Institute of High Energy Physics, Chinese Academy of Sciences, Beijing 100049}
\affiliation{Institute of High Energy Physics, Vienna 1050}
\affiliation{Institute for High Energy Physics, Protvino 142281}
\affiliation{INFN - Sezione di Napoli, 80126 Napoli}
\affiliation{INFN - Sezione di Torino, 10125 Torino}
\affiliation{Advanced Science Research Center, Japan Atomic Energy Agency, Naka 319-1195}
\affiliation{J. Stefan Institute, 1000 Ljubljana}
\affiliation{Kanagawa University, Yokohama 221-8686}
\affiliation{Institut f\"ur Experimentelle Kernphysik, Karlsruher Institut f\"ur Technologie, 76131 Karlsruhe}
\affiliation{Kennesaw State University, Kennesaw, Georgia 30144}
\affiliation{King Abdulaziz City for Science and Technology, Riyadh 11442}
\affiliation{Department of Physics, Faculty of Science, King Abdulaziz University, Jeddah 21589}
\affiliation{Korea Institute of Science and Technology Information, Daejeon 305-806}
\affiliation{Korea University, Seoul 136-713}
\affiliation{Kyoto University, Kyoto 606-8502}
\affiliation{Kyungpook National University, Daegu 702-701}
\affiliation{\'Ecole Polytechnique F\'ed\'erale de Lausanne (EPFL), Lausanne 1015}
\affiliation{P.N. Lebedev Physical Institute of the Russian Academy of Sciences, Moscow 119991}
\affiliation{Ludwig Maximilians University, 80539 Munich}
\affiliation{Luther College, Decorah, Iowa 52101}
\affiliation{University of Maribor, 2000 Maribor}
\affiliation{Max-Planck-Institut f\"ur Physik, 80805 M\"unchen}
\affiliation{School of Physics, University of Melbourne, Victoria 3010}
\affiliation{Moscow Physical Engineering Institute, Moscow 115409}
\affiliation{Moscow Institute of Physics and Technology, Moscow Region 141700}
\affiliation{Graduate School of Science, Nagoya University, Nagoya 464-8602}
\affiliation{Nara Women's University, Nara 630-8506}
\affiliation{National Central University, Chung-li 32054}
\affiliation{National United University, Miao Li 36003}
\affiliation{Department of Physics, National Taiwan University, Taipei 10617}
\affiliation{H. Niewodniczanski Institute of Nuclear Physics, Krakow 31-342}
\affiliation{Nippon Dental University, Niigata 951-8580}
\affiliation{Niigata University, Niigata 950-2181}
\affiliation{Novosibirsk State University, Novosibirsk 630090}
\affiliation{Osaka City University, Osaka 558-8585}
\affiliation{Pacific Northwest National Laboratory, Richland, Washington 99352}
\affiliation{University of Pittsburgh, Pittsburgh, Pennsylvania 15260}
\affiliation{Theoretical Research Division, Nishina Center, RIKEN, Saitama 351-0198}
\affiliation{University of Science and Technology of China, Hefei 230026}
\affiliation{Showa Pharmaceutical University, Tokyo 194-8543}
\affiliation{Soongsil University, Seoul 156-743}
\affiliation{Stefan Meyer Institute for Subatomic Physics, Vienna 1090}
\affiliation{Sungkyunkwan University, Suwon 440-746}
\affiliation{School of Physics, University of Sydney, New South Wales 2006}
\affiliation{Department of Physics, Faculty of Science, University of Tabuk, Tabuk 71451}
\affiliation{Tata Institute of Fundamental Research, Mumbai 400005}
\affiliation{Excellence Cluster Universe, Technische Universit\"at M\"unchen, 85748 Garching}
\affiliation{Department of Physics, Technische Universit\"at M\"unchen, 85748 Garching}
\affiliation{Toho University, Funabashi 274-8510}
\affiliation{Department of Physics, Tohoku University, Sendai 980-8578}
\affiliation{Department of Physics, University of Tokyo, Tokyo 113-0033}
\affiliation{Tokyo Institute of Technology, Tokyo 152-8550}
\affiliation{Tokyo Metropolitan University, Tokyo 192-0397}
\affiliation{University of Torino, 10124 Torino}
\affiliation{Virginia Polytechnic Institute and State University, Blacksburg, Virginia 24061}
\affiliation{Wayne State University, Detroit, Michigan 48202}
\affiliation{Yamagata University, Yamagata 990-8560}
\affiliation{Yonsei University, Seoul 120-749}
 \author{B.~Pal}\affiliation{University of Cincinnati, Cincinnati, Ohio 45221} 
 \author{A.~J.~Schwartz}\affiliation{University of Cincinnati, Cincinnati, Ohio 45221} 
  \author{I.~Adachi}\affiliation{High Energy Accelerator Research Organization (KEK), Tsukuba 305-0801}\affiliation{SOKENDAI (The Graduate University for Advanced Studies), Hayama 240-0193} 
  \author{H.~Aihara}\affiliation{Department of Physics, University of Tokyo, Tokyo 113-0033} 
  \author{S.~Al~Said}\affiliation{Department of Physics, Faculty of Science, University of Tabuk, Tabuk 71451}\affiliation{Department of Physics, Faculty of Science, King Abdulaziz University, Jeddah 21589} 
  \author{D.~M.~Asner}\affiliation{Pacific Northwest National Laboratory, Richland, Washington 99352} 
 \author{T.~Aushev}\affiliation{Moscow Institute of Physics and Technology, Moscow Region 141700} 
 \author{R.~Ayad}\affiliation{Department of Physics, Faculty of Science, University of Tabuk, Tabuk 71451} 
  \author{I.~Badhrees}\affiliation{Department of Physics, Faculty of Science, University of Tabuk, Tabuk 71451}\affiliation{King Abdulaziz City for Science and Technology, Riyadh 11442} 
  \author{A.~M.~Bakich}\affiliation{School of Physics, University of Sydney, New South Wales 2006} 
% ...
\collaboration{The Belle Collaboration}

\begin{abstract}
\noindent
We have searched for the Cabibbo-suppressed decay $\Lambda_c^+\to\phi p\pi^0$ in $e^+e^-$ collisions using a data sample corresponding to
an integrated luminosity of 915 $\rm fb^{-1}$. The data were collected by the Belle experiment  at the 
KEKB $e^+e^-$ asymmetric-energy collider running at or near the $\Upsilon(4S)$ and $\Upsilon(5S)$ resonances. No significant signal is observed, and we set an upper limit on the branching fraction of $\mathcal{B}(\Lambda_c^+\to \phi p\pi^0) <15.3\times10^{-5}$ at 90\% confidence level. 
The contribution of nonresonant $\Lambda_c^+\to K^+K^- p\pi^0$ decays is found to be consistent with zero, and the corresponding upper limit on its  branching fraction is set to be $\mathcal{B}(\Lambda_c^+\to K^+K^-p\pi^0)_{\rm NR} <6.3\times10^{-5} $ at 90\% confidence level. We also search for an intermediate 
hidden-strangeness pentaquark decay $P^+_s\to\phi p$. 
We see no evidence for this intermediate decay and set 
an upper limit on the product branching fraction of
${\cal B}(\Lambda_c^+\to P^+_s \pi^0)\times {\cal B}(P^+_s\to\phi p)
<8.3\times 10^{-5}$ at 90\% confidence level. Finally, 
we  measure the branching fraction for the Cabibbo-favored decay $\Lambda_c^+\to K^-\pi^+p\pi^0$;  the result is  $\mathcal{B}(\Lambda_c^+\to K^-\pi^+p\pi^0)= (4.42\pm0.05\, (\rm stat.) \pm 0.12\, (\rm syst.) \pm 0.16\, (norm.))\%$, which is the most precise measurement to date.
\end{abstract}


\maketitle

\begin{figure}[htb]
\centering
\includegraphics[width=0.245\textwidth]{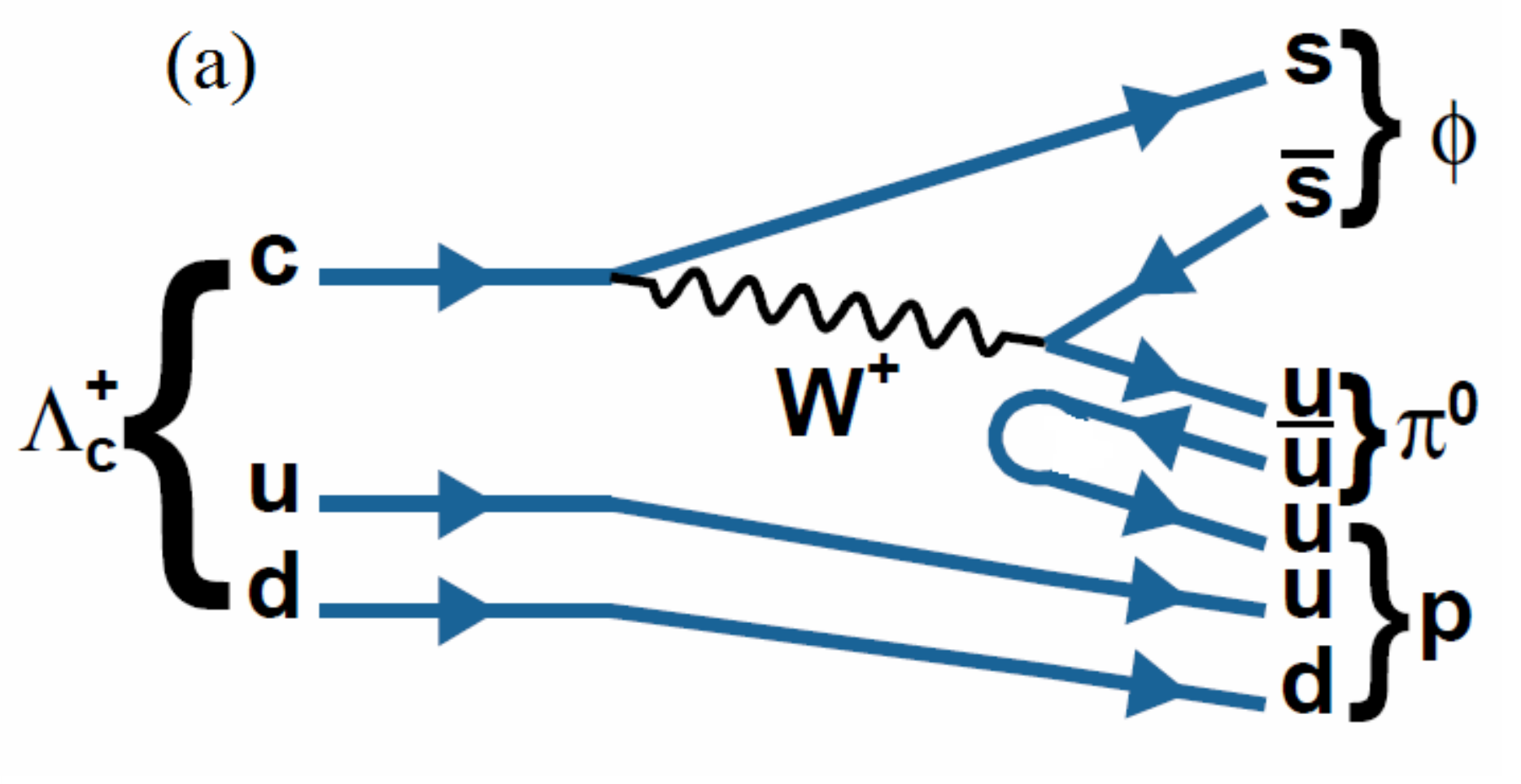}%
\includegraphics[width=0.245\textwidth]{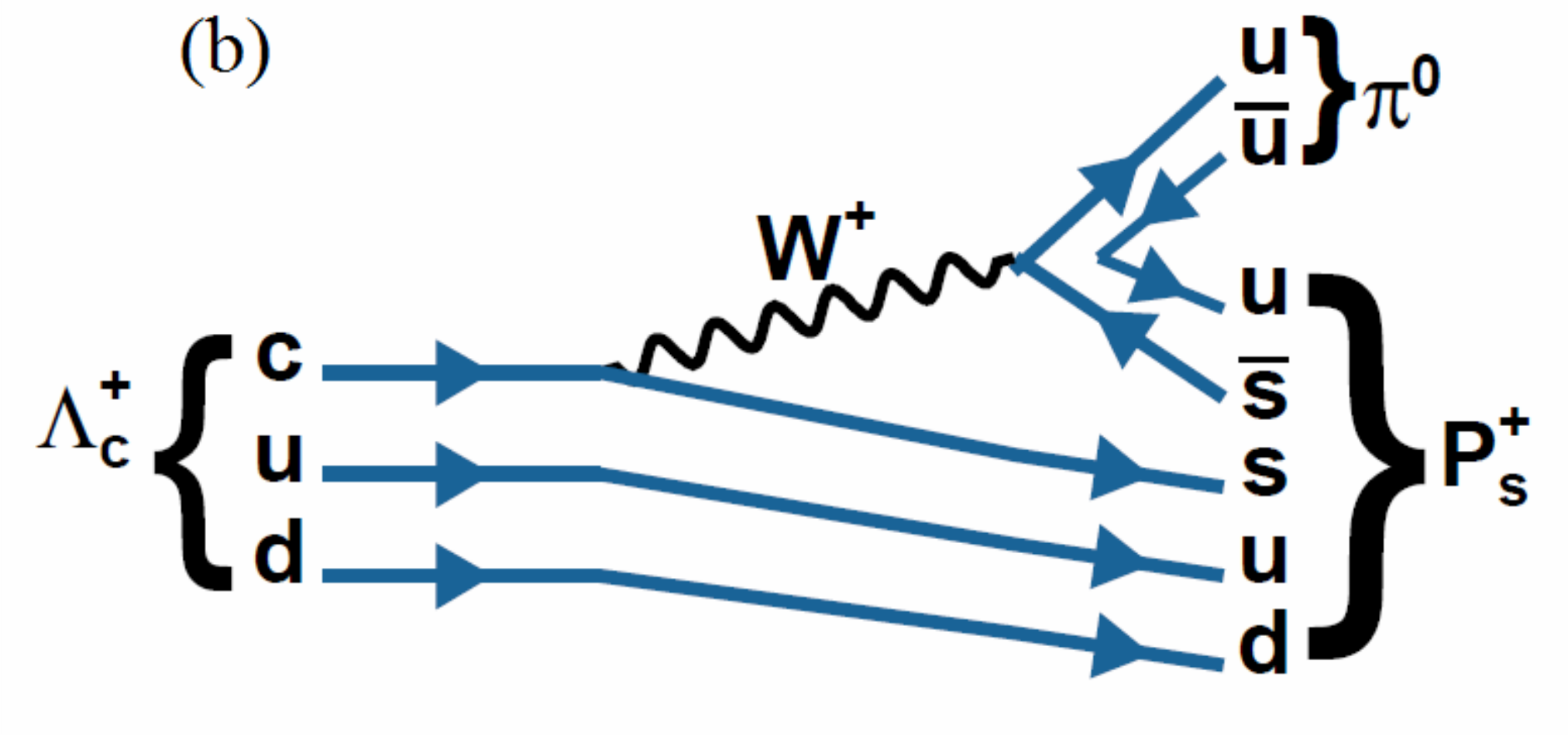}
\caption{\small Feynman diagram for the decay (a) $\Lambda_c^+\to\phi p\pi^0$ and (b) $\Lambda_c^+\to P_s^+\pi^0$.}
\label{fig:Feynman}
\end{figure}

\tighten
The story of exotic hadron  spectroscopy   begins  with the  discovery  of  the
$X(3872)$ by the Belle collaboration in 2003~\cite{Choi:2003ue}. Since then, many exotic $X\!Y\!Z$ states have been reported by Belle and other experiments~\cite{Rev_xyz}. Recent observations of two hidden-charm pentaquark  states $P_c^+(4380)$ and $P_c^+(4450)$ by the LHCb collaboration in the $J/\psi p$ invariant mass spectrum of the  $\Lambda_b^0\to J/\psi pK^- $ process~\cite{Aaij:2015tga} raises the question of whether a hidden-strangeness pentaquark  $P_s^+$, where the $c\bar{c}$ pair  in  $P_c^+$  is replaced by  an $s\bar{s}$ pair, exists~\cite{Kopeliovich:2015vqa, Zhu:2015bba, Lebed:2015dca}. The strange-flavor analogue of the $P_c^+$ discovery channel is the decay $\Lambda_c^+\to\phi p\pi^0$~\cite{Kopeliovich:2015vqa, Lebed:2015dca}, shown in Fig.~\ref{fig:Feynman} (a)~\cite{charge-conjugate}. The detection of a hidden-strangeness pentaquark could be possible through  the $\phi p$ invariant mass spectrum within this channel [see Fig.~\ref{fig:Feynman} (b)],
if the underlying mechanism creating the $P_c^+$ states also holds for $P_s^+$, independent of the flavor~\cite{Lebed:2015dca}, and only if  the mass of $P_s^+$ is less than $M_{\Lambda_c^+}-M_{\pi^0}$.  
In an analogous $s\bar{s}$ process of $\phi$ photoproduction $(\gamma p\to\phi p)$,  a forward-angle  bump structure at $\sqrt{s}\approx2.0$ GeV 
has been observed by  the LEPS~\cite{Mibe:2005er} and CLAS collaborations~\cite{Dey:2014tfa}.
However, this structure appears only at the most forward angles, 
which is  not   expected for  the decay of a resonance~\cite{Lebed:2015fpa}.

Previously, the decay $\Lambda_c^+\to\phi p\pi^0$ has not been studied by any experiment. In this paper, we report a search for this decay  using a data set corresponding to an integrated luminosity of 915 $\rm fb^{-1}$ collected with the Belle detector~\cite{Belle} recorded at or near the $\Upsilon(4S)$ and $\Upsilon(5S)$ resonances at the KEKB asymmetric-energy $e^+e^-$ (3.5 on 8.0~GeV) collider~\cite{KEKB}. 
In addition, we search for the nonresonant decay $\Lambda_c^+\to K^+K^-p\pi^0$ and measure the branching fraction of the  Cabibbo-favored decay $\Lambda_c^+\to K^-\pi^+p\pi^0$.

The Belle detector is 
described in detail elsewhere~\cite{Belle}. To calculate the detector acceptance and reconstruction efficiencies and to study background, we use Monte Carlo (MC) simulated events. The MC events are generated uniformly in phase space  with
{\mbox{\textsc{EvtGen}}\xspace}~\cite{Lange:2001uf} and {\mbox{\textsc{JetSet}}\xspace}~\cite{Sjostrand:1993yb}; the detector response is modeled using {\mbox{\textsc{Geant3}}\xspace}~\cite{geant3}. Final-state radiation is taken into account using the {\mbox{\textsc{Photos}}\xspace}~\cite{Barberio:1993qi} package.

The reconstruction of   $\Lambda_c^+\to \phi p\pi^0$ (and  $\Lambda_c^+\to K^- \pi^+ p\pi^0$) decays proceeds by first reconstructing $\pi^0\to\gamma\gamma$ candidates. 
An electromagnetic calorimeter (ECL) cluster not matched to any track is identified as a photon candidate. Such candidates are required to have an energy greater than 50~MeV in the barrel region and 100~MeV in the endcap regions, where the barrel region covers the polar angle range $32^{\circ} < \theta < 130^{\circ}$, and the endcap regions cover the ranges $12^{\circ} < \theta < 32^{\circ}$ and $130^{\circ}<\theta <157^{\circ}$.  To reject showers produced by neutral hadrons, the photon energy deposited in the $3\times3$ array of ECL crystals centered on the crystal with the highest energy must exceed 80\% of the energy deposited in the corresponding $5\times5$ array of crystals.  
We require that the $\gamma\gamma$ invariant mass be within  0.020 GeV/$c^2$ (about $3.5\sigma$ in resolution)  of  the known $\pi^0$ mass~\cite{PDG}.
To improve the $\pi^0$ momentum resolution, we perform a mass-constrained fit and require that the resulting $\chi^2$ be less than 30. In addition, the momentum of the $\pi^0$ candidates in the center-of-mass (CM) frame is required to be higher than 0.30 GeV/$c$.

We subsequently combine $\pi^0$ candidates with three charged tracks.
Such tracks  are identified using  requirements on the distance of closest approach with respect to the interaction point  along the $z$ axis (antiparallel to the $e^+$ beam) of $|dz|< 1.0$ cm, and in the transverse plane of $dr<0.1$  cm. In addition, charged tracks are required to have a minimum number of hits in the vertex detector ($>1$ in both  the  $z$ and transverse directions). Information obtained  from the central drift chamber, the time-of-flight scintillation counters, and the aerogel threshold Cherenkov counters is combined  to form a likelihood $\mathcal{L}$ for hadron identification. A charged track with the likelihood ratios of $\mathcal{L}_K/(\mathcal{L}_{\pi}+\mathcal{L}_K)> 0.9$ and $\mathcal{L}_K/(\mathcal{L}_{p}+\mathcal{L}_K) > 0.6$;  $\mathcal{L}_K/(\mathcal{L}_{\pi}+\mathcal{L}_K)<0.6$ and $\mathcal{L}_{\pi}/(\mathcal{L}_{p}+\mathcal{L}_{\pi}) > 0.6$; and  $\mathcal{L}_p/(\mathcal{L}_{p}+\mathcal{L}_K) > 0.9$ and $\mathcal{L}_p/(\mathcal{L}_{p}+\mathcal{L}_{\pi})> 0.9$ is regarded as kaon, pion and  proton, respectively. The  efficiencies of these requirements for kaons, pions, and protons are 77\%, 97\%, and 75\%, respectively. The probabilities for a kaon, pion, or proton to be misidentified  are $\mathcal{P}(K\to\pi)\approx10\%$, $\mathcal{P}(K\to p)\approx1\%$; $\mathcal{P}(\pi\to K)\approx1\%$, $\mathcal{P}(\pi\to p)<1\%$; and $\mathcal{P}(p\to K)\approx7\%$, $\mathcal{P}(p\to \pi)\approx1\%$.  Candidate $\phi$ mesons are formed from two oppositely charged tracks that have been identified as kaons. 
We accept events in the wide $K^+K^-$ mass range  $m(K^+K^-)\in(0.99,~1.13)$~GeV/$c^2$.
To suppress combinatorial background, especially from $B$ meson decays, we require that the scaled momentum $(x_p=Pc/\sqrt{E^2_{\rm CM}/4-M^2c^4})$  be greater than 0.45, where $E_{\rm CM}$ is the total CM energy, and $P$ and $M$ are the momentum and invariant mass of the $\Lambda_c^+$ candidates. 
A vertex fit is performed to the charged tracks to form a $\Lambda_c^+$ vertex, and we require that the $\chi^2$ from the fit be less than 50. The decay $\Lambda_c^+\to\Sigma^+\phi$ has the same final state as  the signal decay and is Cabibbo-favored. To avoid contamination from this decay, we reject candidates in which the $p\pi^0$ system has an invariant mass within 0.010 GeV/$c^2$ of the known $\Sigma^+$ mass~\cite{PDG}.
We extract the $\Lambda_c^+$ yield in a signal region that spans  $2.5\sigma$ in resolution 
around the $\Lambda_c^+$  mass~\cite{PDG}; this range corresponds to $\pm0.015$~GeV/$c^2$ for $\Lambda_c\to K^-\pi^+p\pi^0$  and approximately $\pm0.010$~GeV/$c^2$ for the other decays studied.

After applying all these selection criteria, about 16\% of events  in the signal region have multiple $\Lambda_c^+$ candidates.
For these events, we retain the candidate having the smallest sum of $\chi^2$ values obtained from the $\pi^0$ mass-constrained fit and the $\Lambda_c^+$ vertex fit. According to MC simulation, this criterion selects the correct $\Lambda_c^+$ candidate in 72\% of multiple-candidate events. 

\begin{figure*}[h!tp]
\begin{center}
    \includegraphics[width=0.495\textwidth]{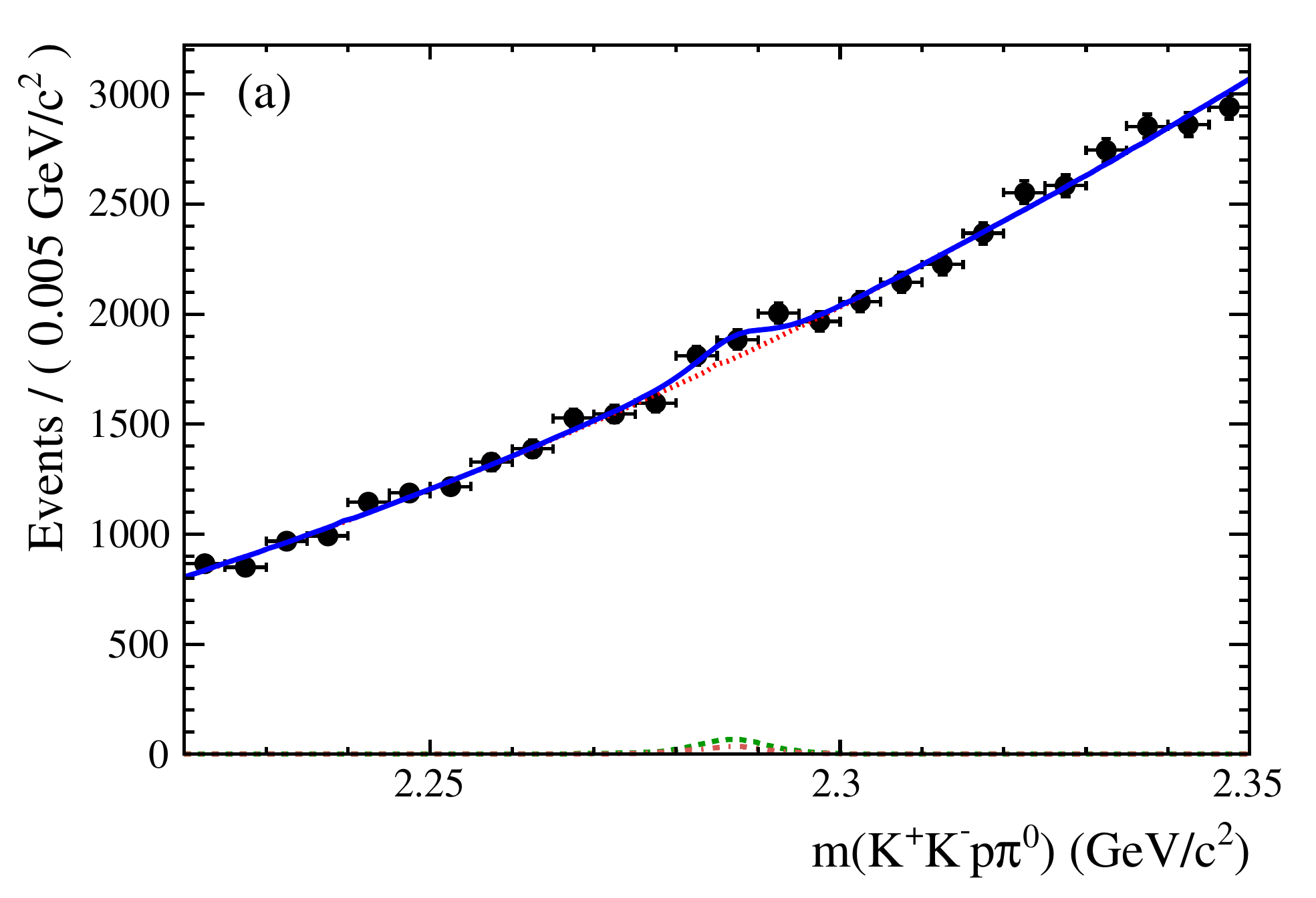}%
       \includegraphics[width=0.495\textwidth]{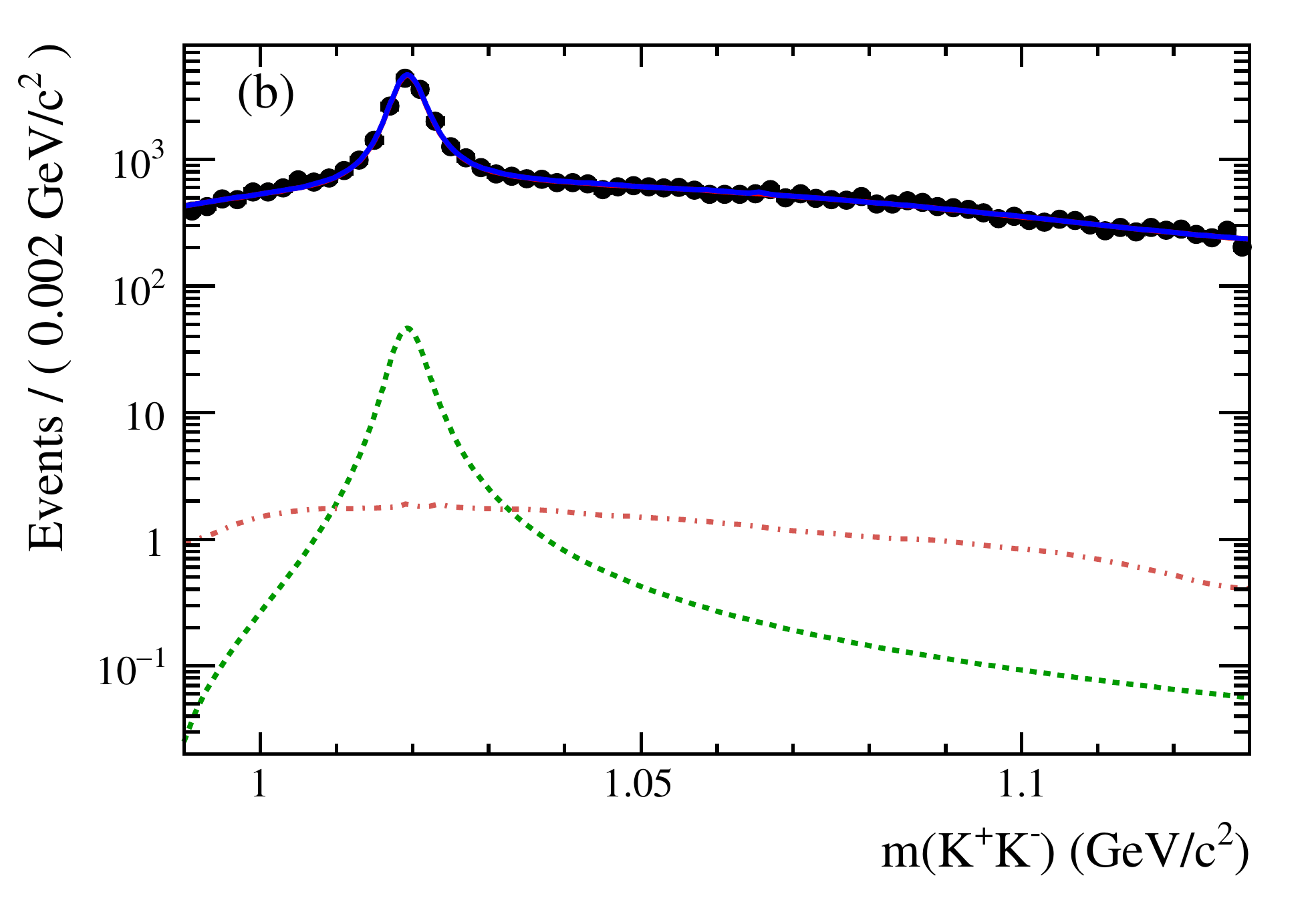}
\end{center}
\vskip -0.5cm
\caption{\small Projections of the 2D fit: (a) $m(K^+K^- p\pi^0)$ and (b) $m(K^+K^-)$. The points with the error bars are the  data, and the (red) dotted, (green) dashed and (brown) dot-dashed curves represent the combinatorial,  signal and nonresonant candidates, respectively, and (blue) solid curves represent the total PDF. The solid curve in (b) completely overlaps the curve for the combinatorial background.}
\label{fig:2dfit}
\end{figure*}

In order to extract the signal yield, we perform a two-dimensional (2D) unbinned extended maximum likelihood fit to the variables $m (K^+K^-p\pi^0)$ and  $m(K^+K^-)$.  Our likelihood function  accounts for three components: $\phi p\pi^0$ signal, $K^+K^-p\pi^0$ nonresonant events, and  combinatorial background. The likelihood  function is defined as
\begin{linenomath}
\begin{equation}
e^{-\sum_j Y_j}\prod_i^N\left( \sum_j Y_j\mathcal{P}_j\big[m^i(K^+K^- p\pi^0),m^i(K^+K^-)\big]\right),
\end{equation}
\end{linenomath}
where $N$ is the total number of events, $\mathcal{P}_j\big[m^i(K^+K^- p\pi^0),m^i(K^+K^-)\big]$  is the probability density function (PDF) of signal or background component $j$ for event $i$, and $j$ runs over all signal and background components. The parameter $Y_j$ is the yield of component $j$. 
The $m(K^+K^-p\pi^0)$ for signal and nonresonant contributions are modeled with the sum of two Crystal Ball (CB) functions~\cite{Skwarnicki:1986xj} having a common mean, whereas for the combinatorial background, a second-order Chebyshev polynomial  is used. The peak positions and resolutions of the CB functions are
adjusted according to data-MC differences observed in the high statistics  sample of $\Lambda_c^+\to K^-\pi^+p\pi^0$ decays. The $m(K^+K^-)$ of signal is modeled with a relativistic Breit-Wigner  function convolved with a Gaussian resolution function ($\rm RBW\otimes G$), with the mass and width of the resonance $\phi$  fixed to their nominal values~\cite{PDG}. The width of the Gaussian resolution function is fixed to the value obtained from the MC simulation. The $m(K^+K^-)$ of nonresonant  background is modeled with a one-dimensional nonparametric PDF~\cite{Cranmer:2000du}.  The $m(K^+K^-)$ of combinatorial background  is modeled with the sum of  a third-order Chebyshev polynomial and the same ${\rm RBW\otimes G}$ function as used to model the signal.
The floated parameters are the component yields $Y_j$ and, for the combinatorial background, the coefficients of the Chebyshev polynomials and the fraction of the RBW. All other parameters are fixed in the fit to the  values obtained from the MC simulation.
Projections of the fit result are shown in Fig.~\ref{fig:2dfit}.
From the fit, we extract $148.4\pm61.8$ signal events, $75.9\pm84.8$ nonresonant events, 
and $7158.4\pm36.4$ combinatorial background events in the $\Lambda_c^+$ signal region.  The statistical significance is evaluated as $\sqrt{-2\ln(\mathcal{L}_0/\mathcal{L}_{\rm max})}$, where $\mathcal{L}_0$ is the likelihood value when the signal yield is fixed to zero, and $\mathcal{L}_{\rm max}$ is  the nominal likelihood value. The statistical significances are found to be 2.4 and 1.0 standard deviations for $\Lambda_c^+\to\phi p \pi^0$ and nonresonant $\Lambda_c^+\to K^+K^- p \pi^0$ decays, respectively.

\begin{figure}[htb]
\centering
\includegraphics[width=0.5\textwidth]{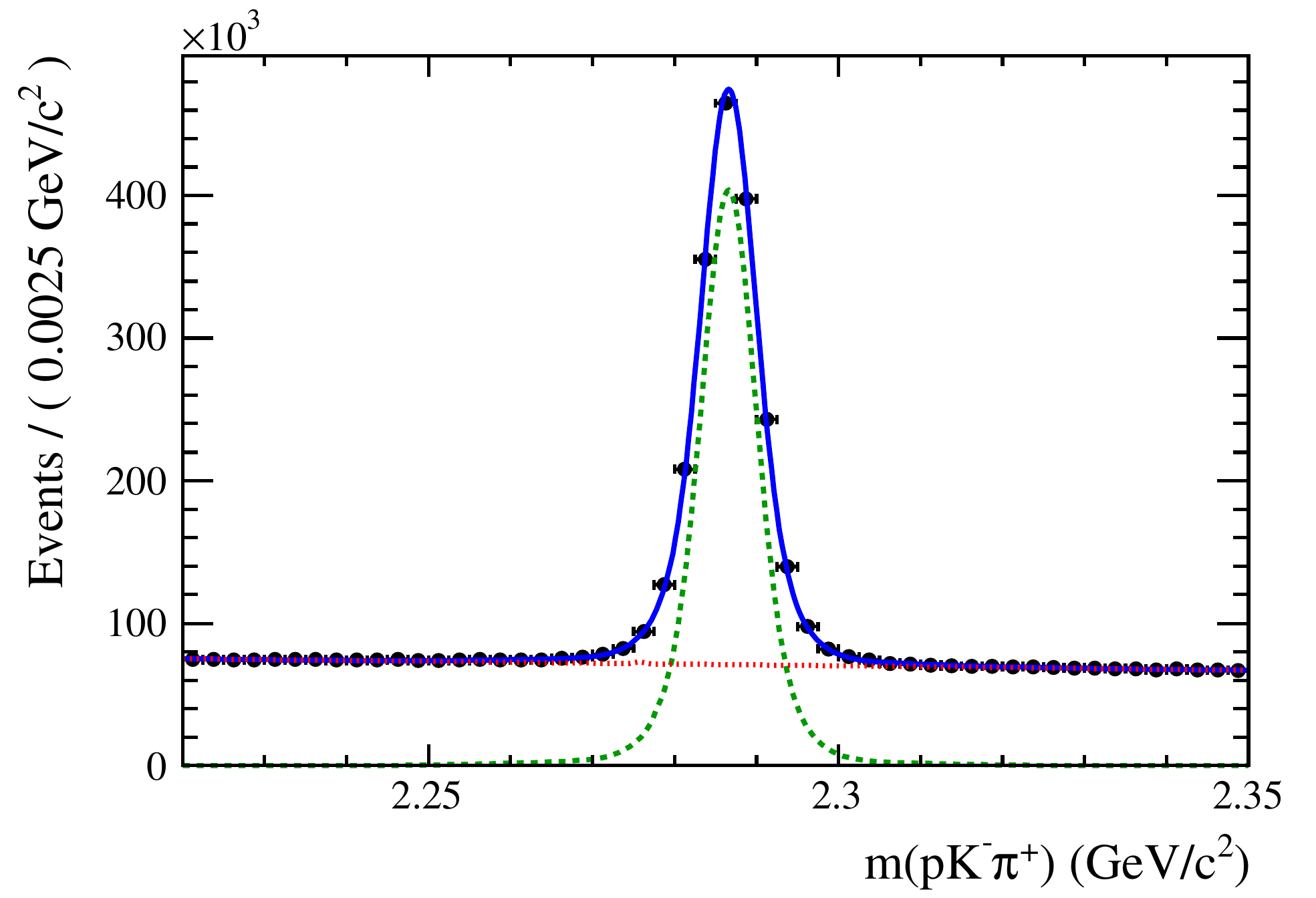}
\caption{\small  Fit to the invariant mass distribution of $p K^-\pi^+$. The points with the error bars are the  data,  the (red) dotted and (green) dashed curves  represent the combinatorial and  signal  candidates, respectively, and (blue) curve represents the total PDF.}
\label{fig:invmass_control_data}
\end{figure}

We use the well-established decay $\Lambda_c^+\to p K^-\pi^+$~\cite{PDG} as the normalization channel for the branching fraction measurements. 
The track, particle identification,  and vertex selection criteria are similar to those used for the signal decays. If there are multiple candidates present in an event, we select the candidate  having the smallest value of $\chi^2$ from the $\Lambda_c^+$ vertex fit.
The resulting invariant mass distribution of the $pK^-\pi^+$ candidates is shown in Fig.~\ref{fig:invmass_control_data}. The signal is modeled with the sum of three Gaussian functions, and the combinatorial background is modeled with a linear function. There are  $1\,468\,435\pm 4816$ signal candidates and $567\,855\pm815$ background candidates in the $\Lambda_c^+$ signal region.

The ratio of branching fractions is calculated as
\begin{linenomath}
\begin{eqnarray}
\frac{\mathcal{B}(\Lambda_c^+\to {\rm final~state})}{\mathcal{B}(\Lambda_c^+\to pK^-\pi^+)} &=& \frac{Y_{\rm Sig}/\varepsilon_{\rm Sig}}{Y_{\rm Norm}/\varepsilon_{\rm Norm}} ,\label{eq:br}
\end{eqnarray}
\end{linenomath}
where $Y$ represents the observed yield in the signal region of the decay of interest and $\varepsilon$ corresponds to the reconstruction efficiency as obtained from the MC simulation. For the $\phi p\pi^0$ final state, we include  $\mathcal{B}(\phi\to K^+K^-)=(48.9\,\pm0.5)\%$~\cite{PDG} in $\varepsilon_{\rm sig}$ of Eq.~(\ref{eq:br}).
The reconstruction efficiencies are $(2.165\,\pm 0.007)\%$, $(2.291\,\pm0.008)\%$, and $(16.564\,\pm0.023)\%$  for $\phi p\pi^0$, nonresonant $K^+K^-p\pi^0$, and  $p K^-\pi^+$ final states, respectively, where the errors are due to MC statistics only.  The ratio $\varepsilon_{\rm Sig}/\varepsilon_{\rm Norm}$  is corrected by a factor $1.028 \pm 0.018$ to  account for   small differences in particle identification efficiencies between data and simulation.  This correction is estimated from a sample of $D^{*+}\to D^0(\to K^-\pi^+)\pi^+$ decays. 
For the $\phi p\pi^0$ final state, the ratio  is 
\begin{linenomath}
\begin{eqnarray}
\frac{\mathcal{B}(\Lambda_c^+\to\phi p\pi^0)}{\mathcal{B}(\Lambda_c^+\to pK^-\pi^+)}=(1.538\pm0.641^{+0.077}_{-0.100})\times10^{-3}.\nonumber 
\end{eqnarray}
\end{linenomath}
Whenever two or more uncertainties are quoted, the first is  statistical and the second is systematic.
Using  $\mathcal{B}(\Lambda_c^+\to pK^-\pi^+)=(6.46\pm0.24)\%$~~\cite{Amhis:2016xyh}, we obtain
\begin{linenomath}
\begin{eqnarray}
\mathcal{B}(\Lambda_c^+\to\phi p\pi^0)=(9.94\pm4.14^{+0.50}_{-0.65}\pm0.37)\times10^{-3},\nonumber 
\end{eqnarray}
\end{linenomath}
where the third uncertainty is that due to the branching fraction $\mathcal{B}(\Lambda_c^+\to p K^-\pi^+)$.

Since the significances are below 3.0 standard deviations for both $\phi p\pi^0$ signal and $K^+K^-p\pi^0$ nonresonant decays, we set upper limits on their branching fractions  at 90\% confidence level (C.L.) using a Bayesian approach. The limit is obtained by integrating the
likelihood function from zero to infinity; the value that
corresponds to 90\% of this total area is taken as the
90\% C.L. upper limit.  We include the systematic uncertainty  in the calculation by
convolving the likelihood distribution with a Gaussian function
whose width is set equal to the total systematic uncertainty. 
The results are
\begin{linenomath}
\begin{eqnarray*}
\mathcal{B}(\Lambda_c^+\to \phi p\pi^0) &<& 15.3\times10^{-5} ,\\
\mathcal{B}(\Lambda_c^+\to K^+K^-p\pi^0)_{\rm NR} &<&6.3\times10^{-5} ,
\end{eqnarray*}
\end{linenomath}
which are the first limits on these branching fractions. 

\begin{figure}[htb]
\centering
\includegraphics[width=0.5\textwidth]{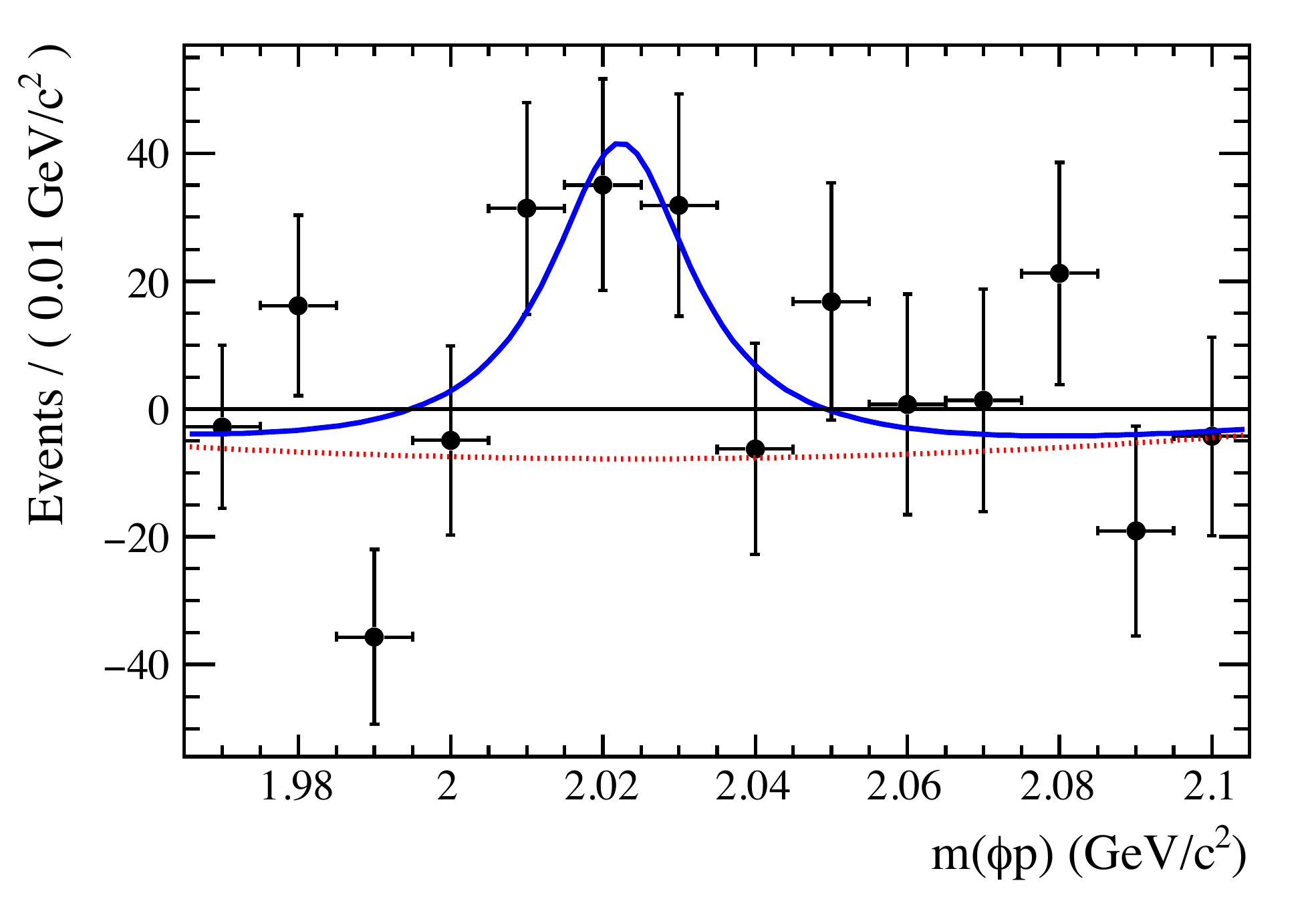}
\caption{\small   The background-subtracted distribution of $m(\phi p)$ in the $\phi p\pi^0$ final state. The points with error bars are data, and   the (blue) solid line shows the total PDF. The (red) dotted curve shows the fitted phase space component (which has fluctuated negative).}
\label{fig:bkg-sub_dis}
\end{figure}

To search for a putative $P_s^+\to\phi p$ decay, we select $\Lambda_c^+\to K^+K^-p\pi^0$ candidates in which $m(K^+K^-)$ is within 0.020~GeV/$c^2$ of the   $\phi$ meson mass~\cite{PDG} 
and plot the  background-subtracted $m(\phi p)$ distribution (Fig.~\ref{fig:bkg-sub_dis}). This distribution is obtained by performing 2D fits as discussed above in bins of $m(\phi p)$. 
The data shows no clear evidence for a $P_s^+$ state. 
We set an upper limit on the product branching fraction 
$\mathcal{B}(\Lambda_c^+\to P_s^+\pi^0) \times \mathcal{B}(P_s^+\to \phi p)$ by fitting the  distribution of Fig.~\ref{fig:bkg-sub_dis}
to the sum of a RBW function and a phase space 
distribution determined from a sample of simulated $\Lambda^+_c\to\phi p\pi^0$ 
decays. We obtain $77.6\pm28.1$ $P_s^+$ events from the fit, which gives an upper limit of 
\begin{eqnarray*}
\mathcal{B}(\Lambda_c^+\to P_s^+\pi^0) \times 
\mathcal{B}(P_s^+\to \phi p) & < & 8.3\times 10^{-5}
\end{eqnarray*}
at 90\% C.L. This limit is calculated using the same procedure as that used for our limit on ${\cal B}(\Lambda_c^+\rightarrow \phi p \pi^0)$. The
systematic uncertainties for the two cases are essentially 
identical except for that due to the size of the MC sample 
used to calculate the reconstruction efficiency.
The  efficiency used here
[$\varepsilon=(2.438\pm 0.026)$\%] 
corresponds to the fitted values
$M_{P_s^+}=(2.025\pm 0.005)$~GeV/$c^2$ and 
$\Gamma_{P_s^+}=(0.022\pm 0.012)$~GeV. 


\begin{figure}[h!tb]
\centering
\includegraphics[width=0.5\textwidth]{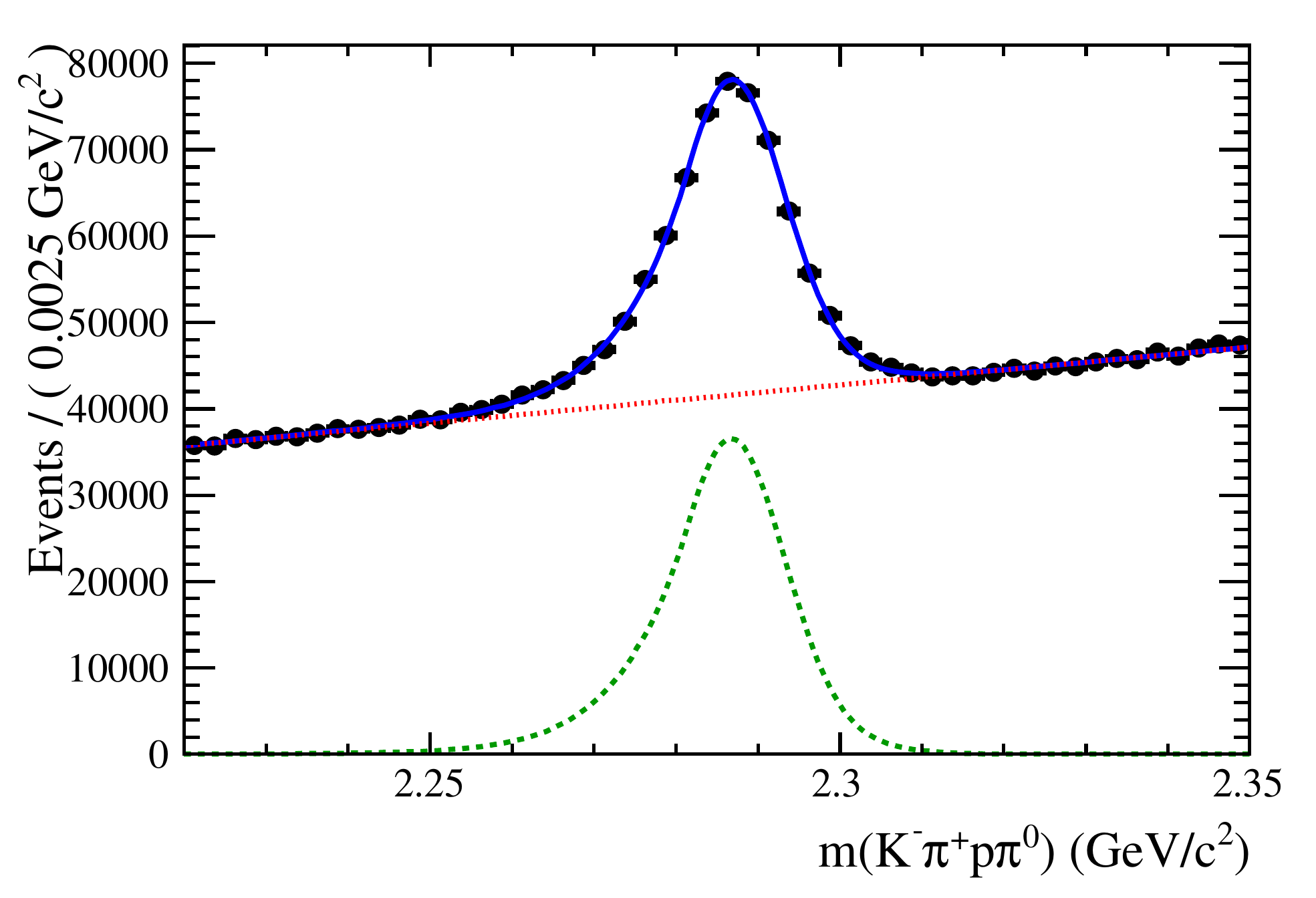}
\caption{\small  Fit to the invariant mass distribution of $m(K^-\pi^+p\pi^0)$. The points with the error bars are the  data,  the (red) dotted and (green) dashed curves  represent the combinatorial and  signal  candidates, respectively, and (blue) curve represents the total PDF.  The $\chi^2/$ (number of bins) of the fit is 1.43, which indicate that the fit gives a good description of the data.}
\label{fig:invmass_control2_data}
\end{figure}

For the $\Lambda_c^+\to K^-\pi^+p\pi^0$ sample, 
the mass distribution is plotted in Fig.~\ref{fig:invmass_control2_data}. We fit this distribution to obtain the signal yield. We model the signal  with a sum of two CB functions having a common mean, and the combinatorial background  with a linear function.
We find $242\,039\pm \,2342$ signal candidates and $472\,729\pm\,467$ background candidates in the $\Lambda_c^+$ signal region.
The corresponding signal efficiency is $(3.988\pm0.009)\%$, obtained from  MC simulation. We measure the ratio of branching fractions
\begin{linenomath}
\begin{equation*}
\frac{\mathcal{B}(\Lambda_c^+\to K^-\pi^+p\pi^0)}{\mathcal{B}(\Lambda_c^+\to K^-\pi^+p)}=(0.685\pm0.007\pm 0.018),
\end{equation*}
\end{linenomath}
which results in a branching fraction
\begin{linenomath}
\begin{equation*}
\mathcal{B}(\Lambda_c^+\to K^-\pi^+p\pi^0)=(4.42\pm0.05\pm 0.12\pm0.16)\%.
\end{equation*}
\end{linenomath}
This is the most precise measurement of $\mathcal{B}(\Lambda_c^+\to K^-\pi^+p\pi^0)$ to date and is consistent with the recently measured value $\mathcal{B}(\Lambda_c^+\to K^-\pi^+p\pi^0)=(4.53\pm0.23\pm0.30)\%$ by the BESIII collaboration~\cite{Ablikim:2015flg}.

The systematic uncertainties on all branching fractions are listed in Table~\ref{sys1}. The uncertainties due to fixed parameters in the PDF shape are estimated by varying the parameters individually according to their statistical uncertainties. For each variation, the branching fraction is recalculated, and the difference with the nominal value is taken as the systematic uncertainty associated with that parameter. In order to determine the systematic uncertainty due to the $m(K^+K^-)$ PDF of nonresonant $K^+K^-p\pi^0$, we  replace the nonparametric PDF by a fourth-order polynomial  and refit the data.  For the $\phi p\pi^0$ final state, we also try including a separate PDF for an $f_0(980)$ intermediate state. The differences in the fit results are included as systematic uncertainties. We add  all uncertainties in quadrature to obtain the overall uncertainty due to PDF parametrization. The uncertainties due to errors in the calibration factors used to account for small data-MC differences in the signal PDF are evaluated separately but in a similar manner. 
\begin{table*}[h!tb]
\renewcommand{\arraystretch}{1.2}
\caption{\small Systematic uncertainties (\%) on $\mathcal{B}(\Lambda_c^+\to \phi p\pi^0)$, $\mathcal{B}(\Lambda_c^+\to K^+K^-p\pi^0)_{\rm NR}$, and $\mathcal{B}(\Lambda_c^+\to K^-\pi^+p\pi^0)$.}
\label{sys1}
\centering
\begin{tabular}{l|cccc}
\hline \hline
Source & $\mathcal{B}(\Lambda_c^+\to \phi p\pi^0)$ & $\mathcal{B}(\Lambda_c^+\to K^+K^-p\pi^0)_{\rm NR}$ & $\mathcal{B}(\Lambda_c^+\to K^-\pi^+p\pi^0)$  \\
\hline
PDF parametrization & $^{+1.0}_{-1.9}$ & $^{+1.9}_{-1.5}$&- \\
Calibration factor &$^{+3.8}_{-5.2}$  & $^{+2.8}_{-1.5}$&-\\
Choice of $m(K^+K^-)$ range & $^{+0.0}_{-1.2}$ &- &-\\
Best candidate selection &2.1 & 2.1 & 2.1  \\
MC sample size &  0.4 & 0.4 & 0.3 \\
$\pi^0$ reconstruction & 1.5 & 1.5 & 1.5 \\
Particle identification & 1.8 & 1.8 & - \\
$\mathcal{B}(\phi\to K^+K^-)$ & 1.0 & - & - \\
\hline
Total (without $\mathcal{B}_{\rm Norm}$) & $^{+5.0}_{-6.5}$ &$^{+4.6}_{-3.8}$ &$2.6$\\
\hline
$\mathcal{B}_{\rm Norm}$ & 3.7 &  3.7 &  3.7 \\
\hline\hline
\end{tabular}
\end{table*}
A  systematic uncertainty of $-1.2\%$ is assigned to account for changes associated with the choice of the $m(K^+K^-)$ range in $\mathcal{B}(\Lambda_c^+\to \phi p\pi^0)$. 
A 2.1\% systematic uncertainty is assigned due to the  best candidate selection. This is evaluated  by analyzing the decay channel $\Lambda_c^+\to\Sigma^+\phi$, which has much higher purity than the signal channels analyzed. 
We determine this by applying an alternative best candidate selection, $i.e.$,  the deviations of the candidate $\phi$ and $\Sigma^+$ masses from their nominal values.  
The difference in the branching fraction due to the two methods of the best candidate selection is taken as the systematic uncertainty.
We assign a 1.5\% systematic uncertainty due to $\pi^0$ reconstruction; this is determined from a study of $\tau^-\to\pi^-\pi^0\nu_{\tau}$ decays. Since the branching fractions are measured with respect to the normalization channel $\Lambda_c^+\to p K^-\pi^+$, which has an identical number of charged tracks, the systematic uncertainty due to differences in  tracking performance between signal and normalization modes is negligible. There is a 1.8\% systematic uncertainty assigned for the  particle identification efficiencies in the $\phi p\pi^0$ and nonresonant $K^+K^- p\pi^0$ final states relative to the $pK^-\pi^+$ normalization channel.  The uncertainty in acceptance due to possible resonance substructure in the decay is found to be negligible. The total of the above systematic uncertainties is calculated as their sum in quadrature. In addition, there is a 3.7\% uncertainty due to the branching fraction of the normalization mode. As this large uncertainty does not arise from our
analysis and  will decrease with future measurements of $\Lambda_c^+\to pK^-\pi^+$, we quote it separately. 

In summary,  we have searched for the decays $\Lambda_c^+\to\phi p\pi^0$ and nonresonant $\Lambda_c^+\to K^+K^- p\pi^0$. No significant signal is observed for either decay mode and we set 90\% C.L. upper limits on their branching fractions, which are
$\mathcal{B}(\Lambda_c^+\to \phi p\pi^0) < 15.3\times10^{-5}$ and 
$\mathcal{B}(\Lambda_c^+\to K^+K^-p\pi^0)_{\rm NR} <6.3\times10^{-5}$.
We see no evidence for a hidden-strangeness pentaquark decay $P_s^+\to \phi p$ and set an upper limit on the product branching fraction of  $\mathcal{B}(\Lambda_c^+\to P_s^+\pi^0) \times \mathcal{B}(P_s^+\to \phi p)<8.3\times10^{-5}$ at 90\% C.L. This limit is a factor of six higher than the product branching fraction measured 
by LHCb for an analogous hidden-charm pentaquark state: 
$\mathcal{B}(\Lambda_b^0\to P_c(4450)^+ K^-) \times 
\mathcal{B}(P_c(4450)^+\to J/\psi\,p)=(1.3\pm 0.4)\times10^{-5}$~\cite{Aaij:2015tga}.
We also measure 
$\mathcal{B}(\Lambda_c^+\to K^-\pi^+p\pi^0)=(4.42\pm0.05\pm 0.12\pm0.16)\%$.
This is the world's most precise measurement of this branching fraction.

\begin{center}
\textbf{ACKNOWLEDGMENTS}
\end{center}
We thank the KEKB group for the excellent operation of the
accelerator; the KEK cryogenics group for the efficient
operation of the solenoid; and the KEK computer group,
the National Institute of Informatics, and the 
PNNL/EMSL computing group for valuable computing
and SINET5 network support.  We acknowledge support from
the Ministry of Education, Culture, Sports, Science, and
Technology (MEXT) of Japan, the Japan Society for the 
Promotion of Science (JSPS), and the Tau-Lepton Physics 
Research Center of Nagoya University; 
the Australian Research Council;
Austrian Science Fund under Grant No.~P 26794-N20;
the National Natural Science Foundation of China under Contracts 
No.~10575109, No.~10775142, No.~10875115, No.~11175187, No.~11475187, 
No.~11521505 and No.~11575017;
the Chinese Academy of Science Center for Excellence in Particle Physics; 
the Ministry of Education, Youth and Sports of the Czech
Republic under Contract No.~LTT17020;
the Carl Zeiss Foundation, the Deutsche Forschungsgemeinschaft, the
Excellence Cluster Universe, and the VolkswagenStiftung;
the Department of Science and Technology of India; 
the Istituto Nazionale di Fisica Nucleare of Italy; 
the WCU program of the Ministry of Education, National Research Foundation (NRF)
of Korea Grants No.~2011-0029457, No.~2012-0008143,
No.~2014R1A2A2A01005286,
No.~2014R1A2A2A01002734, No.~2015R1A2A2A01003280,
No.~2015H1A2A1033649, No.~2016R1D1A1B01010135, No.~2016K1A3A7A09005603, No.~2016K1A3A7A09005604, No.~2016R1D1A1B02012900,
No.~2016K1A3A7A09005606, No.~NRF-2013K1A3A7A06056592;
the Brain Korea 21-Plus program, Radiation Science Research Institute, Foreign Large-size Research Facility Application Supporting project and the Global Science Experimental Data Hub Center of the Korea Institute of Science and Technology Information;
the Polish Ministry of Science and Higher Education and 
the National Science Center;
the Ministry of Education and Science of the Russian Federation and
the Russian Foundation for Basic Research;
the Slovenian Research Agency;
Ikerbasque, Basque Foundation for Science and
MINECO (Juan de la Cierva), Spain;
the Swiss National Science Foundation; 
the Ministry of Education and the Ministry of Science and Technology of Taiwan;
and the U.S.\ Department of Energy and the National Science Foundation.


\begin{thebibliography}{99}
\bibitem{charge-conjugate}
Unless stated otherwise, charge-conjugate modes are implicitly included.
\end{thebibliography}
\end{document}